\documentclass[12pt]{article}

\usepackage[UKenglish]{babel}
\usepackage{latexsym}
\usepackage{amssymb}
\usepackage{epsfig}
\usepackage[colorlinks,hyperindex]{hyperref}

\begin{document}

\title{Recovery of normal heat conduction in harmonic chains with correlated
disorder}

\author{I. F. Herrera-Gonz\'{a}lez${}^{1}$,
F. M. Izrailev${}^{1}$,\\
L. Tessieri${}^{2}$ \\
{\it ${}^{1}$ Instituto de F\'{\i}sica, Universidad Aut\'{o}noma
de Puebla,} \\
{\it Puebla, 72570, Mexico} \\
{\it ${}^{2}$ Instituto de F\'{\i}sica y Matem\'{a}ticas} \\
{\it Universidad Michoacana de San Nicol\'{a}s de Hidalgo} \\
{\it 58060, Morelia, Mexico}}

\date{22nd March 2015}

\maketitle

\begin{abstract}
We consider heat transport in one-dimensional harmonic chains with
isotopic disorder, focussing our attention mainly on how disorder
correlations affect heat conduction.
Our approach reveals that long-range correlations can change the number of
low-frequency extended states. As a result, with a proper choice of
correlations one can control how the conductivity $\kappa$ scales with
the chain length $N$. We present a detailed analysis of the role of specific
long-range correlations for which a size-independent conductivity is exactly
recovered in the case of fixed boundary conditions.
As for free boundary conditions, we show that disorder correlations can lead
to a conductivity scaling as $\kappa \sim N^{\varepsilon}$, with the scaling
exponent $\varepsilon$ being arbitrarily small (although not strictly zero),
so that normal conduction is almost recovered even in this case.
\end{abstract}

Pacs: 44.10.+i, 63.50.Gh, 63.22.-m

\maketitle

\section{Introduction}

The derivation of the transport properties of one-dimensional (1D) models
from the underlying microscopic dynamics is a long-standing problem in
non-equilibrium statistical mechanics. An especially relevant (and
vexed) question is: What conditions are required for heat conduction
to obey Fourier's law (see, e.g.,~\cite{Bon00})?
Among the various 1D models, harmonic chains are the most convenient
for analytical treatment and have been intensively studied over many
decades. The protracted research efforts have led to the conclusion that
Fourier's law does not hold in 1D harmonic chains with isotopic uncorrelated
disorder~\cite{Bon00, Lep03, Dha08}, unless the baths are endowed with
special spectral properties~\cite{Dha01}.
The interest in this topic was significantly heightened after experiments
provided evidence of anomalous heat conduction in carbon and boron-nitride
nanotubes~\cite{Cha04} (see also~\cite{Liu12} for a review of recent
experimental works).

As shown by Matsuda and Ishii~\cite{Mat70}, in finite harmonic chains with
uncorrelated isotopic disorder most normal modes undergo Anderson
localisation and practically do not contribute to heat transport.
The situation is different for low-frequency modes, because their
localisation length exceeds the size of the chain; these modes are
essentially extended and carry energy across the system.
In a random chain of $N$ sites there are about $\sqrt{N}$
low-frequency delocalised modes~\cite{Ish73}; their form depends
crucially on the boundary conditions.
Taking into account these facts, one can compute the heat flux with the
help of the Matsuda-Ishii formula. After dividing the result by the
temperature gradient, one eventually obtains that the conductivity scales
as
\begin{equation}
\kappa \propto N^{\alpha},
\label{power_law_scaling}
\end{equation}
with $\alpha = 1/2$ for free boundary conditions and $\alpha = -1/2$ for
fixed boundary conditions~\cite{Mat70,Cas71,Ver79}.
The conductivity coefficient $\kappa$, therefore,  depends on the size of
the system: this implies that Fourier's law cannot hold, because its
validity rests on $\kappa$ being an intensive quantity.

This phenomenon, known as anomalous heat transport, occurs in random
harmonic chains with {\em uncorrelated} isotopic disorder.
Significant changes of the behaviour of the thermal conductivity can
be expected, however, when disorder exhibits spatial correlations.
In fact, as made clear by the intensive research of the past fifteen years
(see~\cite{Izr12} and references therein), the localisation properties of
the eigenstates of low-dimensional systems can be strongly modified by
imposing specific correlations to the disorder.
Most studies of 1D models with correlated disorder have been focussed on
the localisation of electronic states and electromagnetic or spin waves;
less attention has been devoted to the effects of correlated
disorder on the heat conduction in harmonic chains.
Among the works centred on this topic, one should mention the analysis
done in~\cite{Mou03}, where the authors numerically studied the diffusion
of a localised energy input in random harmonic chains with long-range
disorder correlations.
In Ref.~\cite{Mou03} the sequence of random masses was obtained using the
spectral method for the generation of a fractal Brownian motion.
The same model was analysed in~\cite{Shi07}, where the structure of
the normal modes and of the phonon spectrum was studied.

Rather than focussing on long-range disorder correlations, in
Ref.~\cite{Ong14} the authors considered the effects of exponentially
decaying correlations. They found that correlations of this
kind lead to greater high-frequency phonon transmittance for short chains
and to an increased thermal resistance for long chains.
The impact of specific long-range correlations was studied in
Ref.~\cite{Her10} where it was shown how to delocalise a finite fraction
of the vibrational modes in high-frequency regions of the phononic spectrum.
In this way one can obtain random chains with conductivity roughly
proportional to the system size, i.e., $\kappa \propto N^{\alpha}$ with
$\alpha \sim 1$, as occurs for harmonic chains without disorder.

The previous works on random chains with correlated disorder did not
establish the conditions required for the onset of normal heat conduction,
but their results suggested that Fourier's law might be recovered by
imposing specific correlations to the isotopic disorder.
In this paper we show that this intuition is indeed correct and we
present an approach which allows us to recover Fourier's law in 1D chains
by controlling how $\kappa$ scales with the chain length $N$.

Our idea goes as follows. Instead of using correlations to delocalise
mid-to-high frequency modes, we now employ them to produce significant
alterations of the fraction $N_{\mathrm{e}}/N$ of low-frequency extended modes.
In this way, by means of appropriate correlations, we can modify the scaling
of the conductivity with the system size. In particular, we show that, for
fixed boundary conditions, specific long-range correlations can produce a
size-independent conductivity, thereby ensuring the validity of Fourier's
law even in 1D harmonic chains.
For free boundary conditions one does not obtain an intensive conductivity,
but other long-range correlations can make the conductivity scale with the
size of the chain as $\kappa \sim N^{\varepsilon}$, with $\varepsilon$ positive
but arbitrarily small.

\section{The model}

We consider a 1D harmonic chain of $N$ atoms with nearest-neighbours
interactions and the same elastic constant $k$ throughout the chain.
The corresponding Hamiltonian has the form
\begin{equation}
H = \sum_{n=1}^{N} \frac{p_{n}^{2}}{2m_{n}} + \sum_{n=1}^{N} \sum_{m=1}^{N}
\frac{1}{2} \mathbf{D}_{nm} q_{n} q_{m} .
\label{hamiltonian}
\end{equation}
Here $p_{n}$ and $q_{n}$ respectively represent the momentum and the
displacement from the equilibrium position of the $n$-th atom of mass
$m_{n}$, while $\mathbf{D}$ stands for the tridiagonal force-constant matrix
with elements
$\mathbf{D}_{nm} = 2k \delta_{n,m} - k \delta_{n,m-1} - k \delta_{n,m+1}$
for $n=2,\ldots,N-1$ and $m=1,\ldots, N$. The first and the last
rows of $\mathbf{D}$ determine the dynamical equation of the atoms at
the extremities of the chain; their forms depend on the chosen boundary
conditions.
In the case of free boundary conditions the extreme atoms interact
only with their single nearest neighbour, so that the first and last rows of
the force matrix are
$\mathbf{D}_{1,m} = k \delta_{1,m} - k \delta_{1,m-1}$ and
$\mathbf{D}_{N,m} = k \delta_{N,m} - k \delta_{N,m+1}$.
For fixed boundary conditions, the outermost atoms of the chain
are coupled with springs of elastic constant $k$ not only to their
nearest neighbours, but also to external walls of infinite mass.
Correspondingly, one has $\mathbf{D}_{1,m} = 2k \delta_{1,m} - k \delta_{1,m-1}$
and $\mathbf{D}_{N,m} = 2k \delta_{N,m} - k \delta_{N,m+1}$.

Isotopic disorder enters the model~(\ref{hamiltonian}) via random atomic
masses $m_{n}$ which fluctuate around a common average value
\begin{equation}
M = \langle m_{n} \rangle.
\label{avmass}
\end{equation}
In Eq.~(\ref{avmass}) and in what follows the symbol
$\langle \cdots \rangle$ denotes the average over realisations of the
isotopic disorder. We use the symbol
\begin{equation}
\delta m_{n} = m_{n} - M
\label{mass_fluct}
\end{equation}
for the fluctuation of the $n$-th mass with respect to the average $M$.
The strength of disorder is characterised by the variance of the
fluctuations
\begin{equation}
\langle \left(\delta m_{n} \right)^{2} \rangle = \sigma^{2} ,
\label{sigma}
\end{equation}
which we assume to be finite.
Another key ingredient is the normalised binary correlator
\begin{equation}
\frac{\langle \delta m_{n} \delta m_{n+l}\rangle}{\sigma^{2}}
= \chi(l) ,
\label{bincor}
\end{equation}
whose specific form will be defined below according to the desidered
properties of the thermal conductivity.
We stress that we do not need to assume the usual weak-disorder
condition, $\sigma^{2} \ll 1$. In the present work we are interested
solely in the behaviour of the low-frequency modes, since only these
modes come into play. For this reason Eqs.~(\ref{sigma}) and~(\ref{bincor})
represent a sufficient description of the statistical properties of the
random masses.

We remark that our model is characterised by isotopic disorder. One can
consider also chains with random spring constants rather than random
masses.
However, the model with random masses is more appropriate for the study
of thermal transport, since the isotopical doping is more feasible
experimentally, see Ref.~\cite{Liu12}. It should also be noted that random
springs endow the model with off-diagonal disorder, whereas random masses
generate purely diagonal disorder, which is mathematically easier to deal
with.

To study heat transfer, the extremities of the chain must be coupled to two
heat baths of given temperatures $T_{-}$ and $T_{+}$.
In the literature several models have been used for the baths. Two examples
are: the Langevin representation, adopted in the Casher-Lebowitz
model~\cite{Cas71}, and the semi-infinite harmonic chain~\cite{Rub71}.
Recently, a generalised Langevin scheme was also considered in
Ref.~\cite{Dha01}, demostrating the role of the spectral properties of the
baths in shaping the transport properties of the chain.
To make our approach more transparent, in this work
we use the time-honoured Langevin representation of the baths.
We would like to stress, however, that our results are equally valid for
oscillators baths, as can be shown with the use of the formalism introduced
by Dhar~\cite{Dha01,Roy08}.

In the Langevin scheme the action of the baths is mimicked by adding
to the dynamical equations of the first and of the last atom an extra
force composed of a dissipative term (proportional to the speed of the atom)
and of a Gaussian white noise $\xi(t)$. The dynamical equations of the chain
thus take the form
\begin{equation}
\begin{array}{ccl}
\dot{q}_{n} & = & \displaystyle \frac{p_{n}}{m_{n}} \\
\dot{p}_{n} & = & \displaystyle
\sum_{m=n-1}^{n+1} {\bf D}_{nm} q_{m} +
\delta_{n1} \left[ \xi_{-}(t) -\lambda \frac{{p}_{1}}{m_{1}} \right] \\
& + & \displaystyle
\delta_{nN} \left[ \xi_{+}(t) -\lambda \frac{{p}_{N}}{m_{N}} \right]
\end{array}
\label{dyneq}
\end{equation}
with $n=1, \ldots, N$ and $q_{0} = q_{N+1} = 0$.
In Eq.~(\ref{dyneq}), $\xi_{+}(t)$ and $\xi_{-}(t)$ represent independent
Gaussian white noises with zero average, $\overline{\xi_{\pm}(t)} = 0$, and
\begin{equation}
\begin{array}{rcl}
\overline{\xi_{+}(t) \xi_{+}(t^{\prime})} & = & 2 \lambda k_{B} T_{+} 
\delta \left(t - t^{\prime} \right) \\
\overline{\xi_{-}(t) \xi_{-}(t^{\prime})} & = & 2 \lambda k_{B} T_{-}
\delta \left(t - t^{\prime} \right) \\
\overline{\xi_{-}(t) \xi_{+}(t^{\prime})}& = & 0 .
\end{array}
\label{noise_corr}
\end{equation}
We use the symbol $\overline{ (\cdots)}$ to represent the average
over different realisations of the thermal noises.
As can be seen from Eqs.~(\ref{dyneq}) and~(\ref{noise_corr}), both baths are
coupled with the same constant $\lambda$ to the chain.

As is well known, in harmonic chains the heat flux is given by the sum of
the modal fluxes. In the stationary regime and for weak coupling of the
chain to the baths, the total flux is given by the Matsuda-Ishii
formula~\cite{Mat70}
\begin{equation}
J = \lambda k_{B} \left( T_{+} - T_{-} \right) \sum_{i=1}^{N}
\frac{ \left[ e_{1}^{(i)} \right]^{2} \left[ e_{N}^{(i)} \right]^{2}}
{m_{N} \left[ e_{1}^{(i)} \right]^{2} + m_{1} \left[ e_{N}^{(i)} \right]^{2}}
\label{mi_formula}
\end{equation}
where the symbols $e_{n}^{(i)}$ represents the $n$-th component of the
$i$-th displacement eigenvector, as defined by the equation
\begin{equation}
\tilde{\mathbf{D}} e^{(i)} =  \omega_{i}^{2} e^{(i)} .
\label{displ_ev}
\end{equation}
In Eq.~(\ref{displ_ev}), $\tilde{\mathbf{D}}$ is the rescaled force-constant
matrix with elements
$\tilde{\mathbf{D}}_{nm} = \mathbf{D}_{nm}/\sqrt{m_{n} m_{m}}$
and the chain eigenfrequencies are labelled in ascending order:
$\omega_{1} < \omega_{2} < \ldots < \omega_{N}$.

\section{Correlated disorder}

The Hamiltonian~(\ref{hamiltonian}) defines the autonomous dynamics of
the chain. Taking the Fourier transform of the dynamical equations with
respect to time, one obtains
\begin{equation}
q_{n+1} + q_{n-1} + 4 \left( \frac{\omega}{\omega_{\mathrm{max}}} \right)^{2}
\frac{\delta m_{n}}{M} \; q_{n} =
\left[ 2 - 4 \left( \frac{\omega}{\omega_{\mathrm{max}}}\right)^{2} \right]
q_{n}
\label{dyneq1}
\end{equation}
where
\begin{equation}
\omega_{\mathrm{max}} = \sqrt{\frac{4k}{M}} 
\end{equation}
is the upper limit of the frequency spectrum for the homogeneous chain.
In the absence of disorder, Eq.~(\ref{dyneq1}) reduces to the form
\begin{equation}
q_{n+1} + q_{n-1} =
\left[ 2 - 4 \left(\frac{\omega}{\omega_{\mathrm{max}}}\right)^{2} \right] q_{n}
\label{nodisorder}
\end{equation}
which has solutions in the form of plane waves
\begin{equation}
q_{n} = C \exp \left( \pm i \mu n \right) 
\end{equation}
with the wave number $\mu$.
Substituting the plane-wave solution in Eq.~(\ref{nodisorder}), one obtains
the dispersion relation for the infinite homogeneous chain
\begin{equation}
\omega \left( \mu \right) =
\omega_{\mathrm{max}} \sqrt{\frac{1 - \cos \mu}{2}} =
\omega_{\mathrm{max}} \left| \sin \frac{\mu}{2} \right|,
\label{dispersion}
\end{equation}
The allowed values of the wavenumber $\mu$ depend on the boundary conditions.
One has
\begin{equation}
\mu_{i} = \left\{ \begin{array}{ccl}
\displaystyle
\frac{\pi i}{N+1} & & \mbox{for fixed boundary conditions} \\
\displaystyle
\frac{\pi (i-1)}{N} & & \mbox{for free boundary conditions} \\
\end{array} \right.
\end{equation}
with $i = 1, \ldots, N$. In what follows we will use the short-hand
notation $\omega_{i} = \omega(\mu_{i})$.

When disorder is added, the normal modes suffer Anderson localisation,
provided that the size of the chain is sufficiently large.
We remark that the effective random potential felt by the $i$-th mode is
given by
\begin{equation}
V_{n}^{(i)} = 4 \left( \frac{\omega_{i}}{\omega_{\mathrm{max}}} \right)^{2}
\frac{\delta m_{n}}{M} .
\label{eff_pot}
\end{equation}
Taking into account Eqs.~(\ref{mass_fluct}) and~(\ref{sigma}), it is
easy to see that the effective potential~(\ref{eff_pot}) has zero average,
$\langle V_{n}^{(i)} \rangle = 0$, and variance
\begin{equation}
\Big\langle \left( V_{n}^{(i)}\right)^{2} \Big\rangle
= 16 \left(\frac{\omega_{i}}{\omega_{\mathrm{max}}}\right)^{4}
\frac{\sigma^{2}}{M^{2}}.
\end{equation}
We restrict our attention to the low-frequency modes, identified by the
condition
\begin{equation}
\omega_{i} \ll \omega_{\mathrm{max}}.
\label{low_freq}
\end{equation}
For these modes the effective random potential is weak, as can be seen
by considering that the deterministic part of the diagonal term in
Eq.~(\ref{dyneq1}) is much larger than the random potential~(\ref{eff_pot}).
In fact, when condition~(\ref{low_freq}) is satisfied, one has
\begin{equation}
\sqrt{\Big\langle \left( V_{n}^{(i)} \right)^{2} \Big\rangle} \ll 
2 - 4 \left( \frac{\omega_{i}}{\omega_{\mathrm{max}}} \right)^{2} .
\label{weak_eff_dis}
\end{equation}
Note that condition~(\ref{low_freq}) ensures that the
relation~(\ref{weak_eff_dis}) holds even if $\sigma/M \sim 1$.

For the low-frequency modes one can use the Hamiltonian map approach
described in~\cite{Her10} and obtain the following expression for the
inverse localisation length
\begin{equation}
l^{-1}_{\mathrm{loc}}(\omega_{i}) \simeq \frac{\sigma_{i}^{2}}{2 M^{2}}
\left( \frac{\omega_{i}}{\omega_{\mathrm{max}}} \right)^{2}
W\left( \frac{2 \omega_{i}}{\omega_{\mathrm{max}}} \right) .
\label{lyap}
\end{equation}
In Eq.~(\ref{lyap}) the factor $W$ represents the power spectrum of the
random sequence
$\{\delta m_{n} \}$, i.e.,
\begin{equation}
W(\mu) = 1 + 2 \sum_{l=1}^{\infty} \chi(l) \cos \left( 2 l \mu \right)
= \sum_{l=-\infty}^{\infty} \chi (l) \exp \left( i 2 \mu l \right) .
\label{ps}
\end{equation}

Following~\cite{Mat70}, we can now use formula~(\ref{lyap}) to estimate
the number $N_{\mathrm e}$ of extended modes. In a finite chain, a mode
can be considered extended if its localisation length $l_{\mathrm{loc}}$
exceeds the chain length $N$, i.e., if the condition
\begin{equation}
N \lesssim l_{\mathrm{loc}}(\omega_{i}) = \frac{2M^{2}}{\sigma^{2}}
\left( \frac{\omega_{\mathrm{max}}}{\omega_{i}} \right)^{2}
\frac{1}{W \left( 2 \omega_{i}/\omega_{\mathrm{max}} \right)}
\label{deloc_cond}
\end{equation}
is satisfied.
Because low-frequency modes experience weak disorder, their eigenfrequencies
stay close to the corresponding values for an homogeneous chain and one can
still use Eq.~(\ref{dispersion}). In the low-frequency region, the
dispersion law~(\ref{dispersion}) can be linearised, so that in the end
one can substitute in Eq.~(\ref{deloc_cond}) the approximate expression
\begin{equation}
\omega_{i} \simeq \omega_{\mathrm{max}} \frac{\pi i}{2N} ,
\label{freq}
\end{equation}
with $i \ll N$.
Condition~(\ref{deloc_cond}), supplemented by Eq.~(\ref{freq}),
leads to the conclusion that the number $N_{\mathrm{e}}$ of extended modes
is obtained by solving the equation
\begin{equation}
N_{\mathrm{e}} \simeq \frac{2 M}{\sigma}
\sqrt{\frac{2N}{W \left( \pi N_{\mathrm{e}}/N \right)}} .
\label{exten_modes}
\end{equation}

\section{Emergence of normal conductivity}

To estimate the heat flux, we now evaluate the sum in Eq.~(\ref{mi_formula})
by taking into account that the main contribution is due to the
$N_{\mathrm e}$ extended modes, and that they have essentially the same
form of those of an ordered chain.
In this way, for $N \gg 1$ one obtains
\begin{equation}
\begin{array}{ccl}
J & \simeq & \displaystyle
\frac{\lambda k_{B} (T_{+} - T_{-})}{M N}
\sum_{k=1}^{N_{e}} \sin^{2} \left( \frac{\pi k}{N+1} \right) \\
& \sim & \displaystyle
\frac{\lambda k_{B} (T_{+} - T_{-})}{M} \left( \frac{N_{\mathrm{e}}}{N} \right)^{3}
\end{array}
\label{jfixed}
\end{equation}
for fixed boundary conditions and
\begin{equation}
J \simeq \frac{\lambda k_{B} (T_{+} - T_{-})}{M N}
\sum_{k=1}^{N_{e}} \cos^{2} \left( \frac{\pi k}{2N} \right) \sim
\frac{\lambda k_{B} (T_{+} - T_{-})}{M} \frac{N_{\mathrm{e}}}{N}
\label{jfree}
\end{equation}
for free boundary conditions.
Dividing the heat flux by the temperature gradient, one eventually
arrives at the result that the conductivity scales as
\begin{equation}
\kappa \sim \left\{ \begin{array}{ccl}
N_{\mathrm{e}}^{3}/N^{2} & & \mbox{for fixed boundary conditions} \\
N_{\mathrm{e}} & & \mbox{for free boundary conditions} \\
\end{array} \right.
\label{kappa_scaling_law}
\end{equation}

For uncorrelated disorder $W = 1$ and with Eq.~(\ref{exten_modes}) we
recover the well-known result~\cite{Ish73}
\begin{equation}
N_{\mathrm{e}} \sim \sqrt{N} .
\end{equation}
Substituting this result in Eq.~(\ref{kappa_scaling_law}) one
obtains that $\kappa \sim 1/\sqrt{N}$  for fixed boundary
conditions and that $\kappa \sim \sqrt{N}$ for free boundary conditions.
These remarkable results were established in the early 1970s by several
authors~\cite{Mat70,Cas71,Rub71,Ver79} and led to the crucial conclusion that
Fourier's law {\em cannot} hold in 1D disordered chains. Since then, this 
discovery has been the subject of an intense debate (see, e.g.,~\cite{Lep03}
and references therein).

The aim of this paper is to analyse whether spatial correlations of the
disorder can produce an intensive conductivity and thus lead to normal
heat conduction. We also want to explore the possibility to have other
scaling laws for $\kappa$ as a function of $N$.
To proceed in this direction one has to understand first how specific
correlations of the isotopic disorder can change the number of low-frequency
extended states and, therefore, alter the scaling behaviour of the
conductivity.
Let us assume that in the long-wavelength limit the power spectrum~(\ref{ps})
obeys a power law of the form
\begin{equation}
\begin{array}{ccc}
W(\mu) \sim \mu^{\beta} & \mbox{ for } & \mu \to 0 .
\end{array}
\label{low_freq_ps}
\end{equation}
Note that here $\beta > -1$; the values $\beta \leq -1$ are forbidden
because the power spectrum~(\ref{ps}) must obey the normalisation condition
\begin{equation}
\int_{0}^{\pi/2} W(\mu) \mathrm{d} \mu = \frac{\pi}{2} .
\label{normalisation}
\end{equation}
Substituting Eq.~(\ref{low_freq_ps}) into the condition~(\ref{deloc_cond}),
one obtains that the number of extended modes becomes
\begin{equation}
N_{e} \sim N^{\gamma}
\label{ext_modes}
\end{equation}
with
\begin{equation}
\gamma = \frac{1 + \beta}{2 + \beta} .
\end{equation}
Since $\beta > -1$, the values of $\gamma$ are restricted to the interval
$0 < \gamma < 1$.
After substituting the result~(\ref{ext_modes}) in
Eq.~(\ref{kappa_scaling_law}), one can conclude that the conductivity obeys
the power law~(\ref{power_law_scaling}) with scaling exponent $\alpha$ equal
to
\begin{equation}
\begin{array}{lcl}
\displaystyle
\alpha = \frac{\beta - 1}{\beta + 2} & &
\mbox{for fixed boundary conditions} \\
\displaystyle
\alpha = \frac{\beta + 1}{\beta + 2}  & &
\mbox{for free boundary conditions} .\\
\end{array}
\label{kappa_scaling}
\end{equation}
Eq.~(\ref{kappa_scaling}) represents the central result of this paper.
It shows how disorder correlations can change the scaling of the conductivity
with the size of the chain.

Two special cases stand out for their interest. For fixed boundary conditions,
one can consider the value $\beta = 1$ in Eq.~(\ref{low_freq_ps}).
This implies that the power spectrum decreases linearly with the frequency
in the limit $\omega \to 0$; the corresponding binary correlator~(\ref{bincor})
decays with a power-law $\chi(l) \sim 1/l^{2}$ for $l \gg 1$.
From Eq.~(\ref{kappa_scaling}), it is easy to see that for $\beta = 1$ the
scaling exponent of the conductivity vanishes, so that the conductivity
becomes independent of the size of the chain, $\kappa \sim N^{0}$.
The physical mechanism behind this result is easy to understand.
For $\beta = 1$, the inverse localisation length~(\ref{lyap}) scales
as $l_{\mathrm loc}^{-1} \sim \omega^{3}$ rather than being a quadratic function
of the frequency as in the case of uncorrelated disorder. Therefore, the
number of extended modes is {\em increased} due to the chosen correlations,
and one has $N_{\mathrm e} \sim N^{2/3}$ instead of $N_{\mathrm e} \sim N^{1/2}$.
The increase of summands in Eq.~(\ref{jfixed}) compensates the $1/N^{2}$
scaling of the displacement eigenvectors and produces an intensive conductivity.
Therefore, Fourier's law emerges in the asymptotic regime, i.e., for
$N \gg 1$.

For free boundary conditions, a case of physical interest corresponds to
a power spectrum {\em diverging} for $\omega \to 0$ as
$W(\omega) \sim 1/\omega^{1-\varepsilon}$. This corresponds to
$\beta = - 1 + \varepsilon$, with $\varepsilon$ being a small but positive
parameter. In this case the correlation function decays extremely slowly
over large distances: $\chi(l) \sim 1/l^{\varepsilon}$ for $l \gg 1$.
The choice $\beta = - 1 + \varepsilon$ implies that the localisation
length scales as $l_{\mathrm loc} \sim \omega^{1 + \varepsilon}$ in
the low-frequency region. Therefore the number of extended eigenmodes is
now {\em reduced} with respect to the case of uncorrelated disorder; in fact,
one has $N_{\mathrm e} \sim N^{\varepsilon}$ for $\varepsilon \to 0^{+}$.
This reduction of the extended modes leads to the result
\begin{equation}
\kappa \sim N^{\varepsilon/(1+\varepsilon)} \sim N^{\varepsilon} .
\end{equation}
Since condition~(\ref{normalisation}) requires $\varepsilon > 0$, one
cannot exactly obtain a normal conduction for free boundary conditions
by manipulating the low-frequency behaviour of the power spectrum.
However, for sufficiently small values of $\varepsilon$, one can obtain
a conductivity with a very weak dependence on the size of the chain.

\begin{figure}[!ht]
\begin{center}
\includegraphics[width=5in,height=3.5in]{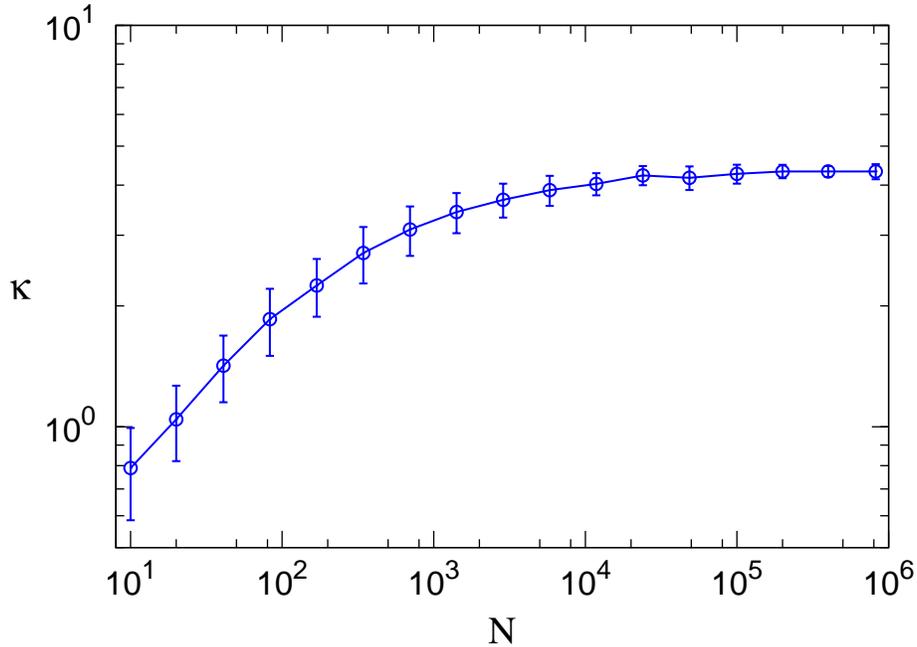}
\caption{\label{kappa_fixed_bc}
Conductivity $\kappa$ versus chain length $N$ for fixed boundary conditions.
Here we considered fluctuations $\delta m_{n}$ of the random masses with the
power spectrum~(\ref{ps_fixed_bc}) which, in the low-frequency
region, behaves as Eq.~(\ref{low_freq_ps}) with $\beta=1$.
The parameters of the model used in simulations were:
$M = 1$, $k = 1$, $\lambda = 1$, and $\sigma^{2} = 0.2$.
The points represent the mean value of the conductivity, obtained after
averaging over 100 disorder realisations. Error bars represent the standard
deviation of the numerical data with respect to the mean value.}
\end{center}
\end{figure}

\begin{figure}[!ht]
\begin{center}
\includegraphics[width=5in,height=3.5in]{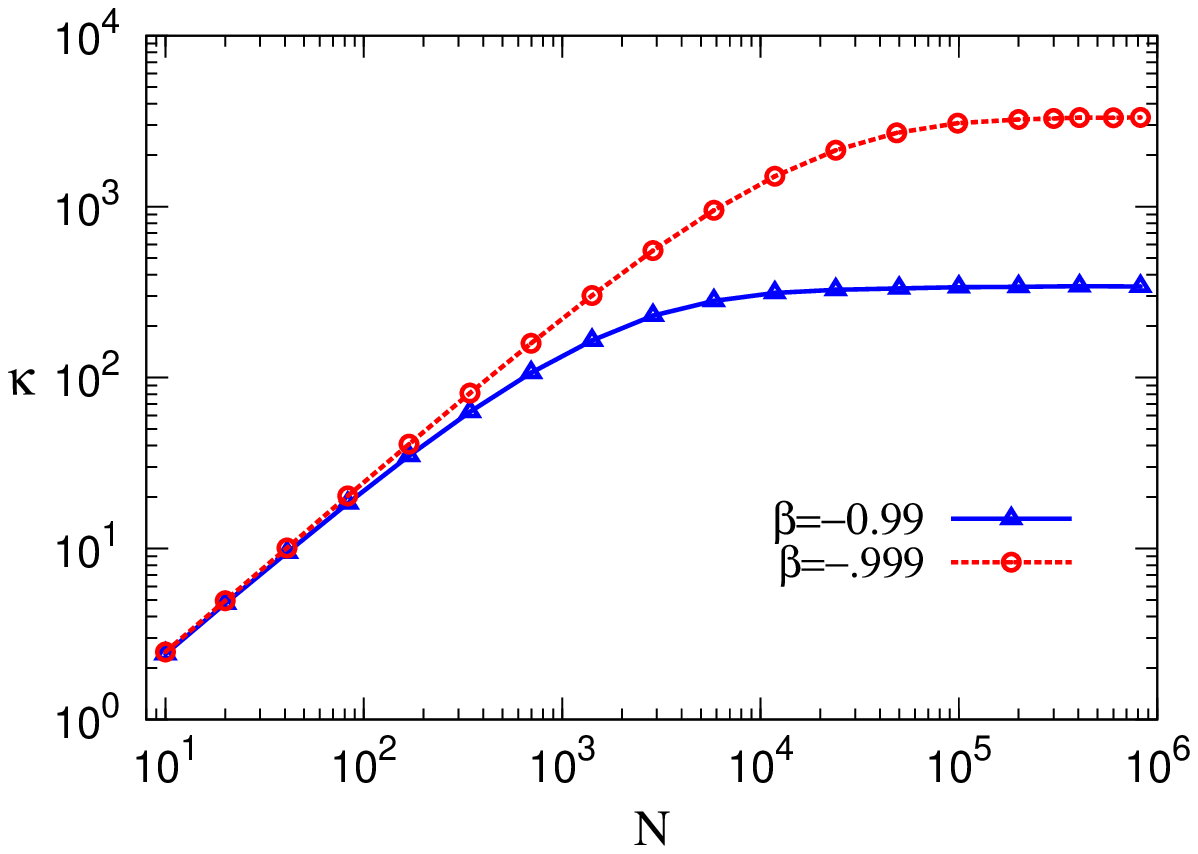}
\caption{\label{kappa_free_bc}
Conductivity $\kappa$ versus chain length $N$ for free boundary conditions.
The fluctuations $\delta m_{n}$ of the random masses have the power
spectrum~(\ref{ps_free_bc}) with $\beta=0.990$ (corresponding
to $\varepsilon = 0.01$) and $\beta = 0.999$ (corresponding to
$\varepsilon = 0.001$). Here the parameters of the model were:
$M = 1$, $k = 1$, $\lambda = 1$, and $\sigma^{2} = 0.2$.
The points represent the mean value of the conductivity, obtained after
averaging over 100 disorder realisations. Error bars are omitted because
they are smaller than the symbols used to represent the mean values of
$\kappa$.}
\end{center}
\end{figure}

These theoretical predictions are confirmed by the numerical data,
as can be seen from Figs.~\ref{kappa_fixed_bc} and~\ref{kappa_free_bc},
which show the scaling behaviour of the conductivity for fixed and
free boundary conditions.
In numerical simulations we set $M = 1$, $k = 1$, $\lambda = 1$, and
$\sigma^{2} = 0.2$. We considered 100 different realisations of the
disorder and we computed the average value and the standard deviation
of the conductivity over this ensemble.
To generate random masses with the desired spatial correlations we have
used the standard technique of filtering white-noise sequences~\cite{Izr12}.

For fixed boundary conditions, we chose a binary correlator of the form
\begin{equation}
\chi(l) = \left( \sqrt{2} + 1 \right) \frac{\sqrt{2} + (-1)^{l+1}}{1 - 16 l^{2}}
\end{equation}
which corresponds to the power spectrum
\begin{equation}
W(\mu) = \frac{\pi (\sqrt{2} + 1)}{2 \sqrt{2}} \sin \left( \frac{\mu}{2}
\right) .
\label{ps_fixed_bc}
\end{equation}
The data represented in Fig.~\ref{kappa_fixed_bc} clearly show how the
conductivity flattens as $N$ increases.

For free boundary conditions we generated a disorder with power spectrum
\begin{equation}
W(\mu) = \frac{1}{1 + \beta} \left( \frac{2}{\pi} \mu \right)^{\beta} =
\frac{1}{\varepsilon} \left( \frac{\pi}{2 \mu} \right)^{1-\varepsilon}
\label{ps_free_bc} 
\end{equation}
with $\varepsilon = 0.01$ and $\varepsilon = 0.001$.
Fig.~\ref{kappa_free_bc} demonstrates that, as $\varepsilon \to 0^{+}$
and $\beta = -1 + \varepsilon$ approaches $-1^{+}$, the asymptotic behaviour
of the conductivity becomes very close to the normal one.
We remark that, as $\beta$ moves closer to $-1$, the size of the ballistic
region becomes larger and larger. This is to be expected, because the random
masses become strongly correlated over very long distances; physically,
this means that the chain is made up of large semi-homogeneous
chunks. As $N$ increases, the almost-normal behaviour is eventually
reached.

\section{Conclusions}

We have analysed the heat conductivity $\kappa$ in harmonic chains with
isotopic correlated disorder. As is known, in such chains normal heat
conduction is not possible when the disorder is uncorrelated. Specifically,
the conductivity $\kappa$, , instead of being independent of the length $N$
of the chain, scales as $\kappa \sim N^\alpha$, where $\alpha$ depends on
the boundary conditions.
Our analytical approach shows that, with a proper choice of long-range
correlations of the disorder, one can recover normal heat conduction for
sufficiently large $N$. The length scale $N_{\mathrm c}$ for the onset of
normal heat conduction, $\kappa \sim N^{0}$ for
$N \gtrsim N_{\mathrm c}$, strongly depends on the boundary conditions.
Specifically, for fixed boundary conditions one can obtain a conductivity
strictly {\em independent} of $N$, while for free boundary conditions normal
heat conduction can be achieved with an arbitrary precision.
With our approach one can further show that, for trickier correlations,
abnormal scaling laws can also emerge, such as $\kappa \sim N/(\ln N)^{\delta}$
with $\delta=1$ for free boundary conditions and $\delta = 3$ for fixed
boundary conditions.
Although such a scaling occurs for extremely large values of $N$, this
example shows that, in principle, one can get a conductivity which increases
with the chain size regardless of the imposed boundary conditions.
In general, our results demonstrate how the behaviour of the heat conductivity
can be controlled by imposing specific correlations in the disorder. This
may prove important in future technological applications.

\section*{Acknowledgments}

The authors acknowledge support from the SEP-CONACYT (M\'{e}xico)
under grant No. CB-2011-01-166382.
I. F. H.-G. and F. M. I. also acknowledge VIEP-BUAP grant MEBJ-EXC12-G and
PIFCA BUAP-CA-169, while L.~T. acknowledges the support of CIC-UMSNH grant
for the years 2014-2015.

\end{document}